\documentclass{llncs}
\usepackage{amssymb}
\usepackage[all]{xy}
\usepackage{ulem}
\newtheorem{prop}[theorem]{Proposition}

\newcommand{\F}{\mbox{\bf  F}}

\def\qed{$\Box$}

\title{Complexity of Decoding Positive-Rate Reed-Solomon Codes}

\author{Qi Cheng\inst{1} and Daqing Wan\inst{2}}

\institute{School of Computer Science\\ 
The University of Oklahoma\\
Norman, OK73019\\
Email: qcheng@cs.ou.edu
\and
Department of Mathematics\\ 
University of California\\
Irvine, CA 92697-3875\\
Email: dwan@math.uci.edu}

\begin{document}

\maketitle

\begin{abstract} The complexity of maximal likelihood decoding
of the  Reed-Solomon codes $[q-1, k]_q$ is a well known open problem. The
only known result \cite{ChengWa07} in this direction states that it
is at least as hard as the discrete logarithm in some cases where
the information rate unfortunately goes to zero. In this paper, we
remove the rate restriction and prove that the same complexity
result holds for any positive information rate. In particular, this
resolves an open problem left in \cite{ChengWa07}, and rules out the
possibility of a polynomial time algorithm for maximal likelihood
decoding problem of Reed-Solomon codes of any rate under a well
known cryptographical hardness assumption. As a side result, we give
an explicit construction of Hamming balls of radius bounded away
from the minimum distance, which contain exponentially many
codewords for Reed-Solomon code of any positive rate less than one.
The previous constructions in \cite{BenKo06}\cite{GuruswamiRu06}
only apply to Reed-Solomon codes of diminishing rates. We also give
an explicit construction of Hamming balls of relative radius less
than $1$ which contain subexponentially many codewords for
Reed-Solomon code of rate approaching one.

\end{abstract}

\setcounter{tocdepth}{1}
\section{Introduction}
Let ${\bf F}_q$ be a finite field of $q$ elements and of characteristic
$p$. A linear error-correcting $[n,k]_q$ code is defined to
be a linear subspace of dimension $k$ in $\F_q^n$.
Let $D=\{x_1,\cdots, x_n\} \subseteq {\bf F}_q$ be a subset of
cardinality $|D|=n>0$. For $1\leq k\leq n$,
let $f$ run over all polynomials in ${\bf F}_q[x]$ of degree at
most $k-1$, the vectors of the form
$$(f(x_1), \cdots, f(x_n))\in {\bf F}_q^n$$
constitute a linear error-correcting $[n,k]_q$ code. If $ D =\F_q^*
$, it is famously known as the  Reed-Solomon
code. If $ D =\F_q $, it is  known as the extended Reed-Solomon
code. We denote them by $RS_q[q-1,k]$ and $RS_q[q,k]$ respectively.
We simply call it a generalized Reed-Solomon code if $D$ is an arbitrary subset
of $\F_q$.

\begin{remark}
In some code theory literature, $RS_q[q-1,k]$
is called primitive Reed-Solomon code, and a generalized Reed-Solomon
code $[n,k]_q $ is defined to be 
$$ \{ (y_1 f(x_1), \cdots, y_n f(x_n)) | f \in \F_q[x], deg(f) < k\},$$
where $ y_1,y_2, \cdots, y_n $ are nonzero elements in  $\F_q$.
\end{remark}

The minimal distance of a generalized Reed-Solomon $[n,k]_q$ code is $n-k+1$
because a non-zero polynomial of degree at most $k-1$ has at most
$k-1$ zeroes. The ultimate decoding problem for an error-correcting
$[n,k]_q$ code is the maximal likelihood decoding: given a received
word $u\in \F_q^n$, find a codeword $v$ such that the Hamming
distance $d(u,v)$ is minimal. When the number of errors is
reasonably small, say, smaller than $n-\sqrt{nk}$, then the list
decoding
algorithms of 
Guruswami-Sudan \cite{GuruswamiSu99} gives a polynomial time
algorithm to find all the codewords for the generalized Reed-Solomon $[n,k]_q$
code.

When the number of errors increases beyond $n-\sqrt{nk}$, it is not
known whether there exists a polynomial time decoding algorithm.
The maximal likelihood decoding of a generalized Reed-Solomon $[n,k]_q$ code is
known to be ${\bf NP}$-complete \cite{GuruswamiVa05a}. The difficulty
is caused by the combinatorial complication of the subset $D$ with
no structures. In fact, there is a straightforward way to reduce the
subset sum problem in $D$ to the deep hole problem of a 
generalized Reed-Solomon
code, which can then be reduced to the maximal likelihood decoding
problem \cite{ChengMu07}.
Note that the subset sum problem for $D\subseteq \F_q$ is hard only if
$|D|$ is much smaller than $q$.

In practical applications, one rarely uses the case of arbitrary
subset $D$. The most widely used case is when $D={\bf F}_q^*$ with
rich algebraic structures. This case is essentially equivalent to
the case $D={\bf F}_q$.  For simplicity,
we focus on the extended Reed-Solomon code $RS_q [q,k]$ in this
paper, all our results can be applied to the  Reed-Solomon
code $RS_q[q-1,k]$ with little modification. The maximal likelihood
decoding problem of $RS_q [q,k]$ is  considered to be hard, but the
attempts to prove its ${\bf NP}$-completeness have failed so far.
The methods in \cite{GuruswamiVa05a}\cite{ChengMu07} can not be
specialized to $RS_q [q,k]$ because we have lost the freedom to
select $D$. The only known complexity result \cite{ChengWa07} in
this direction says that the decoding of $RS_q[q,k]$ is at least as
hard as the discrete logarithm in ${\bf F}_{q^h}^*$ for $h$
satisfying
$$ h \leq  \sqrt{q}-k,
 h\leq  q^{1 \over 2+\epsilon} + 1
{\rm \ \ and\ \ } h\leq { k - { 4 \over \epsilon} - 2
\over {{4\over \epsilon}+1 } }$$
for any $\epsilon > 0 $. The main weakness of this result is that
$\sqrt{q} $ has to be greater than $k$, which implies that the
information rate $k/q$ goes to zero. But in the real world, we tend
to use the Reed-Solomon codes of high rates. Our main result of this
paper is to remove this restriction. Precisely, we show that

\begin{theorem}\label{main}
For any $c \in [0,1]$, there exists an infinite explicit family of
Reed-Solomon codes
$$\{  RS_{q_1}[q_1,k_1], RS_{q_2}[q_2,k_2], \cdots,
RS_{q_i}[q_i,k_i], \cdots \}$$ with $q_i =\Theta ( i^2\log^2 i)$ and $k_i
= (c + o(1)) q_i $ such that if there is a polynomial time
randomized algorithm solving the maximal likelihood decoding problem
for the above family of codes, then there is a polynomial time
randomized algorithm solving the discrete logarithm problem over all
the fields in $\{{\bf F}_{q_1^{h_1}}, {\bf F}_{q_2^{h_2}}, \cdots,
{\bf F}_{q_i^{h_i}}, \cdots \}$, where $h_i $ is any integer less
than $ q_i ^{1/4+o(1)}$.
\end{theorem}


The discrete logarithm problem over finite fields is well studied in
computational number theory. It is not believed to have a polynomial
time algorithm. Many cryptographical protocols base their security
on this assumption. The fastest general purpose algorithm
\cite{JouxLe06} solves the discrete logarithm problem over finite
field $\F^*_{q^h}$ in conjectured time
$$ exp(O( (\log q^h)^{1/3} ( \log\log q^h)^{2/3})). $$
Thus, in
the above theorem, it is best to take $h_i$ as large as possible
(close to $ q_i ^{1/4+o(1)}$) in order for the discrete logarithm to
be hard.
If $h = q^{1/4+o(1)}$, this complexity is subexponential on $ q$.
The above theorem rules out a polynomial time algorithm for the
maximal likelihood decoding problem of Reed-Solomon code of any rate
under a cryptographical hardness assumption.

Our earlier  paper \cite{ChengWa07} proved the theorem 
for $c=0$ (in that case we have
$ h_i \leq q_i ^{1/2+o(1)}$ ).  In this paper, we shall be
concentrating on $0<c\leq 1$. The results in this paper are built on 
the methods and results of our earlier paper.
We shall show that the case $c=1$
follows from the case $c=0$ by a dual argument. The main new idea for the
case $0<c<1$ is to exploit the role of
subfields contained in ${\bf F}_q$. Assume that $q = \tilde{q}^2$
and $h = q^{1/4 + o(1)}$ is a positive integer. We have ${\bf
F}_{\tilde{q}} \subseteq {\bf F}_q \subseteq {\bf F}_{q^h}$. Let
$\alpha $ be an element in ${\bf F}_{q^h}$ such that ${\bf
F}_{\tilde{q}}[\alpha] = {\bf F}_q [\alpha] = {\bf F}_{q^h}$.  We
observe that if every element in $ {\bf F}_{q^h} $ can be written as
a product of $g_1$ many distinct $\alpha + a$ with $a\in {\bf
F}_{\tilde{q}} $, then for any nonnegative integer $ g_2 \leq
q-\tilde{q}$, every element in $ {\bf F}_{q^h} $ can be written as a
product of $g_1+g_2$ many distinct $\alpha + a$
with $a\in {\bf F}_{q} $. 
This observation enables us to prove the
main technical lemma that for any constant $0 < c < 1$, any
element in $\F_{q^h}$ can be written as a product of 
$\lfloor cq \rfloor $ distinct
factors in $\{ \alpha + a | a\in \F_q\}$ for $q$ large enough.


By a direct counting argument, for any positive integer $r<q-k$,
there exists a Hamming ball of radius $r$ containing at least $ { q
\choose r} / q^{q-r-k} $ many codewords in Reed-Solomon code $
RS_q[q,k] $. Thus, if $k= \lfloor cq \rfloor $ for a constant
$0<c<1$, we set $ r= \lfloor q-k - q^{1/4}\rfloor $ and the number
of code words in the Hamming ball will be exponential in $q$.
However, finding such a Hamming ball deterministically is a hard
problem. There are some work done on this problem
\cite{GuruswamiRu06}\cite{BenKo06}, but all the results are for codes of
diminishing rates. Our contribution to this problem is to remove the
rate restriction.

\begin{theorem}\label{manycodewords}
For any $c \in (0,1)$, there exists a deterministic algorithm that
given a positive integer $i$, outputs a prime power $q$, a positive
integer $k$ and a vector $v \in \F_q^q $ such that
\begin{itemize}
\item $q =\Theta ( i^2\log^2 i)$ and $k = (c + o(1)) q $, and
\item the Hamming ball centered at $v$ and of radius
 $q - k - q^{1/4 + o(1)}$
contains  $exp(\Omega(q))$ many codewords in $RS_q[q,k]$, and
\item the algorithm runs in time $i^{O(1)}$.
\end{itemize}
\end{theorem}

In our construction, the ratio between the Hamming ball radius $q -
k - q^{1/4 + o(1)}$ and the minimum distance $q-k+1$, which is known
as the relative radius of the Hamming ball, is approaching $1$. The
same problem was encountered in \cite{GuruswamiRu06}\cite{BenKo06}, where
there is the further restriction that the information rate goes to
zero. In contrast, the above theorem allows the information rate to
be positive. The following result shows that we can decrease the
relative radius to a constant less than $1$ if we work with codes
with information rate going to one.

\begin{theorem}\label{relativeradiuslessthanone}
For any real number $  \rho \in (2/3, 1) $, there is a deterministic
algorithm that, given a positive integer $i$, outputs a prime power
$q = i^{O(1)}$, a positive integer $k = q - o(\sqrt{q})$ and a vector $v \in
\F_q^q$ such that the Hamming ball centered at $v$ and of radius $
[\rho (q -k +1)] $ contains at least $q^i$ many codewords in
$RS_q[q,k]$. The algorithm has time complexity $i^{O(1)}$. Note that
the information rate is $1-o(1)$.
\end{theorem}

It would be interesting for future research to extend the
result to  all $\rho \in  (1/2,1)$, and
to prove a similar result with the
information rate positive and the relative radius less than $1$.

Given a real number $\rho\in (0,1)$, the codes where some Hamming
ball of relative radius $\rho$ contains superpolynomially many
codewords are called $\rho$-dense. It was known in \cite{DumerMi03}
how to efficiently construct such codes for any $\rho \in ( 1/2,
1)$, but finding the center of such a Hamming ball in deterministic
polynomial time is an open problem. 
In this paper, we solve this
problem if the relative radius falls in the range $ (2/3,1) $ using
Reed-Solomon codes  of rate approaching one.
This result derandomizes an important step in  the inapproximability result for
minimum distance problem of a linear code in  \cite{DumerMi03}. 
To completely derandomize the reduction there, however, one needs to
find a linear map from a dense Hamming ball into a linear subspace.
This is again an interesting future research direction.

\section{Previous work for rate $c=0$}

For reader's convenience, in this section, we sketch the main ideas
in our earlier paper \cite{ChengWa07}. This will be the starting
point of our new results in the present paper.

Let $h\geq 2$ be a positive integer. Let $h(x)$ be a monic
irreducible polynomial in ${\bf F}_q[x]$ of degree $h$. Let $\alpha$
be a root of $h(x)$ in an extension field. Then, ${\bf
F}_q[\alpha]={\bf F}_{q^h}$ is a finite field of $q^h$ element. We
have

\begin{theorem}\label{reduction}
Let $h<g<q$ be positive integers. If every element of ${\bf
F}_{q^h}^*$ can be written as a product of exactly $g$ distinct
linear factors of the form $\alpha+a$ with $a\in {\bf F}_q$, then
the discrete logarithm in ${\bf F}_{q^h}^*$ can be efficiently
reduced in random time $q^{O(1)}$ to the maximal likelihood decoding
of the  Reed-Solomon code $RS_q[q, g-h]$.
\end{theorem}
{\bf Proof}.  In \cite{ChengWa07}, the same result was stated for
the weaker bounded distance decoding. Since the specific words used
in \cite{ChengWa07} have exact distance $q-g$ to the code $RS_q[q,
g-h]$, the bounded distance decoding and the maximal likelihood
decoding are equivalent for those special words. Thus, we may
replace bounded distance decoding by the maximal likelihood decoding
in the above statement. We now sketch the main ideas.

Let $h(x)$ be a monic irreducible polynomial of degree $h$ in ${\bf
F}_q[x]$. We shall identify the extension field ${\bf F}_{q^h}$ with
the residue field ${\bf F}_q[x]/(h(x))$. Let $\alpha$ be the class
of $x$ in ${\bf F}_q[x]/(h(x))$. Then, ${\bf F}_q[\alpha] ={\bf
F}_{q^h}$. Consider the  Reed-Solomon code $RS_q[q, g-h]$. For
a polynomial $f(x)\in {\bf F}_q[x]$ of degree at most $h-1$, let
$u_f$ be the received word
$$u_f = ({f(a)\over h(a)} + a^{g-h})_{a\in {\bf F}_q}.$$
By assumption, we can write
$$f(\alpha) =\prod_{i=1}^g (\alpha + a_i),$$
where $a_i\in {\bf F}_q$ are distinct. It follows that as
polynomials, we have the identity
$$\prod_{i=1}^g (x+a_i) = f(x) +t(x)h(x),$$
where $t(x)\in {\bf F}_q[x]$ is some monic polynomial of degree
$g-h$. Thus,
$${f(x)\over h(x)} + x^{g-h} +(t(x)-x^{g-h}) = {\prod_{i=1}^g (x+a_i)\over
h(x)},$$ where $t(x)-x^{g-h}\in {\bf F}_q[x]$ is a polynomial of
degree at most $g-h-1$ and thus corresponds to a codeword. This
equation implies that the distance of the received word $u_f$ to the
code $RS_q[q,g-h]$ is at most $q-g$. If the distance is smaller than
$q-g$, then one gets a monic polynomial of degree $g$ with more than
$g$ distinct roots. Thus, the distance of $u_f$ to the code is
exactly $q-g$.

Let $C_f$ be the set of codewords in $RS_q[q,g-h]$ which has
distance exactly $q-g$ to the received word $u_f$. The cardinality
of $C_f$ is then equal to ${1\over g!}$ times the number of ordered
ways that $f(\alpha)$ can be written as a product of exactly $g$
distinct linear factors of the form $\alpha +a$ with $a\in {\bf
F}_q$. For error radius $q-g$, the maximal likelihood decoding of
the received word $u_f$ is the same as finding a solution to the
equation
$$f(\alpha) =\prod_{i=1}^g (\alpha + a_i),$$
where $a_i\in {\bf F}_q$ being distinct.

To show that the discrete logarithm in ${\bf F}_{q^h}^*$ can be
reduced to the decoding of the words of the type $u_f$, we apply the
index calculus algorithm. Let $b(\alpha)$ be a primitive element of
${\bf F}_{q^h}^*$. Taking $f(\alpha)=b(\alpha)^i$ for a random
$0\leq i\leq q^h-2$, the maximal likelihood decoding of the word
$u_f$ gives a relation
$$b(\alpha)^i = \prod_{j=1}^g (\alpha +a_j(i)),$$
where $a_j(i)\in {\bf F}_q$ are distinct for $1\leq j\leq g$. This
gives the congruence equation
$$i \equiv \sum_{j=1}^g \log_{b(\alpha)}(\alpha +a_j(i))~({\rm mod}~q^h-1).$$
Repeating the decoding and let $i$ vary, this would give enough
linear equations in the $q$ variables $\log_{b(\alpha)}(\alpha
+a)$ ($a\in {\bf F}_q)$). Solving the linear system modulo
$q^h-1$, one finds the values of $\log_{b(\alpha)}(\alpha +a)$ for
all $a\in {\bf F}_q$. To compute the discrete logarithm of an
element $v(\alpha)\in {\bf F}_{q^h}^*$ with respect to the base
$b(\alpha)$, one applies the decoding to the element $v(\alpha)$
 and finds a relation
$$v(\alpha) =\prod_{j=1}^g (\alpha +b_j),$$
where the $b_j\in {\bf F}_q$ are distinct. Then,
$$\log_{b(\alpha)}v(\alpha) \equiv \sum_{j=1}^g
\log_{b(\alpha)}(\alpha +b_j) ~({\rm mod}~ q^h-1).$$ In this way,
the discrete logarithm of $v(\alpha)$ is computed. The detailed
analysis can be found in \cite{ChengWa07}. \hfill\qed

The above theorem is the starting point of our method. In order to
use it, one needs to get good information on the integer $g$
satisfying the assumption of the theorem. This is a difficult
theoretical problem in general. It can be done in some cases, with
the help of Weil's character sum estimate  together with a simple
sieving. Precisely, the following result was proved for $g$ in
\cite{ChengWa07}.

\begin{theorem}\label{gissmall}
Let $h<g$ be positive integers. Let
$$
N(g,h)= {1 \over g!}\left({q^g -{g\choose 2}q^{g-1} \over q^h-1}
-(1+{g\choose 2})(h-1)^g q^{g/2} \right).
$$
Then every element in ${\bf F}_{q^h}^*$ can be written in at least
$N(g,h)$ ways as a product of exactly $g$ distinct linear factors of
the form $\alpha+a$ with $a\in {\bf F}_q$.

If for some constant $\epsilon>0$, we have
$$q\geq \max(g^2, (h-1)^{2+\epsilon}), \ \ g\geq ({4\over
\epsilon}+2)(h+1),$$ then
$$N(g,h)\geq q^{g/2}/g!>0.$$
\end{theorem}

The main draw back of the above theorem is the condition $q\geq
g^2$ which translates to the condition that the information rate
$(g-h)/q$ goes to zero in applications.

\section{The result for rate $c=1$}

Now we show that Theorem~\ref{main} holds when information rate approaches one.

\begin{prop}
Let $g, h$ be positive integers such that for some constant
$\epsilon>0$, we have
$$q\geq \max(g^2, (h-1)^{2+\epsilon}), \ \ g\geq ({4\over
\epsilon}+2)(h+1).$$ Then, every element in ${\bf F}_{q^h}^*$ can
be written in at least $N(g,h)$ ways as a product of exactly $q - g$
distinct linear factors of
the form $\alpha+a$ with $a\in {\bf F}_q$.
\end{prop}

To prove this proposition, we observe that the map that sends
$\beta \in {\bf F}_{q^h}^*$ to $ \prod_{a\in {\bf F}_q} (\alpha + a) /\beta $
is one-to-one from $ {\bf F}_{q^h}^*$ to itself.

{\bf Proof:} Note that
$$ \prod_{a\in {\bf F}_q} (\alpha + a) \not= 0. $$
Given an element $\beta \in {\bf F}_{q^h}^* $, from
Theorem~\ref{gissmall}, we have that $ \prod_{a\in {\bf F}_q}
(\alpha + a) /\beta  $ can be written in at least $N(g,h)$ ways as a
product of exactly $g$ distinct linear factors of the form
$\alpha+a$ with $a\in {\bf F}_q$, hence $\beta$ can be written in at
least $N(g,h)$ ways as a product of exactly $q - g$ distinct linear
factors of the form $\alpha+a$ with $a\in {\bf F}_q$. \hfill\qed

It follows from Theorem~\ref{reduction} that we have the following
two results.

\begin{prop}
Suppose that $$q\geq \max(g^2, (h-1)^{2+\epsilon}), \ \ g\geq
({4\over \epsilon}+2)(h+1).$$ Then the maximal likelihood decoding
$RS_q[q, q-g-h] $ is as hard as the discrete logarithm over the
finite field ${\bf F}_{q^h}^* $.
\end{prop}

Note that the rate $(q-g-h)/q$ approaches $1$ as $q$ increases for
$g=O(\sqrt{q})$ and $h = O(g)=O(\sqrt{q})$.

\begin{prop}
Suppose that $$q\geq \max(g^2, (h-1)^{2+\epsilon}), \ \ g\geq
({4\over \epsilon}+2)(h+1).$$ Let $h(x)$ be an irreducible
polynomial of degree $h$ over $\F_q$ and let $f(x)$ be a nonzero
polynomial of degree less than $h$ over $\F_q$. Then in Reed-Solomon
code $RS_q[q, q-g-h] $, the Hamming ball centered at $({f(a)\over
h(a)} + a^{q-g-h})_{a\in {\bf F}_q}$ of radius $g $ contains at least
$ q^{ g/2} \over g!  $ many codewords.
\end{prop}

Note if we set $g = \lceil \sqrt{q}\rceil $, then the number of
codewords is greater than $2^{\sqrt{q}}$, which is subexponential.

{\bf Proof of Theorem~\ref{relativeradiuslessthanone}:}  The
relative radius of the Hamming ball in the above proposition is $g
\over g+h+1 $. If $g = \lceil ({4\over \epsilon}+2)(h+1) \rceil $,
then the relative radius is approaching to ${4\over \epsilon}+2
\over {4\over \epsilon}+ 3   $ $ =  $ $ 2\epsilon + 4 \over
3\epsilon + 4       $. Select $\epsilon$ such that
$$ \rho = { 2\epsilon + 4 \over  3\epsilon + 4      }. $$
Note that $\epsilon $ can be large if $\rho$ is close to $2/3$.
If $ g = \lceil q^{ 1 \over 2 + \epsilon} \rceil $,
the number of codewords is at least
$$ {q^{ g/2} \over g!}  > (\sqrt{q}/g)^{g}
= q^{ \epsilon g  \over { 2(2+\epsilon)   }  }.$$ To make sure that
this number is greater than $q^i$, we need $g >   {   2(2+\epsilon)i
\over  \epsilon   }      $. It is satisfied if we let $q$ to be the
least prime power which is greater than
$$  ( {2(2+\epsilon)i \over  \epsilon} )  ^{ 2+ \epsilon  } = i^{O(1)}.   $$
We then calculate $g = \lceil q^{ 1 \over 2 + \epsilon} \rceil $ and
solve $h$ from the equation $ g = \lceil ({2\over \epsilon}+2)(h+1)
\rceil    $. Finally we find an irreducible polynomial $h(x)$ of
degree $h$ over $\F_q$ using the algorithm in \cite{Shoup90a}.
\hfill\qed

\section{The result for rate $0<c<1$}

We now consider the positive rate case with $0<c<1$. For this
purpose, we take $q=q_1^m$ with $m\geq 2$. Let $\alpha$ be an
element in $\F_{q^h}$ with $\F_{q_1}[\alpha]  = \F_{q^h}$. Since
$$\F_{q_1}[\alpha] \subseteq \F_q [\alpha] \subseteq F_{q^h},$$
we also have $F_{q^h}= \F_q [\alpha]$.

\begin{theorem}
Let $q=q_1^m$ with $m\geq 2$. Let $g_1$ and $g_2$ be non-negative
integers with $g_2 \leq q-q_1$. Let
$$
N(g_1, g_2, h,m)= {1 \over g_1!} \left({q_1^{g_1} -{g_1\choose
2}q_1^{g_1-1} \over q_1^{mh}-1} -(1+{g_1\choose 2})(mh-1)^{g_1}
q_1^{g_1/2} \right) {q-q_1 \choose g_2 }
$$
Then, every element in ${\bf F}_{q^h}^*$ can be written in at least
$N(g_1, g_2, h,m)$ ways as a product of exactly $g_1+g_2$ distinct
linear factors of the form $\alpha+a$ with $a\in {\bf F}_q$.

If for some constant $\epsilon>0$, we have
$$q_1\geq \max(g_1^2, (mh-1)^{2+\epsilon}), \ \ g_1\geq ({4\over
\epsilon}+2)(mh+1)$$ then
$$N(g_1, g_2, h,m)\geq {q_1^{g_1 /2 } \over g_1! }{q-q_1 \choose
g_2 }>0.$$

\end{theorem}

{\bf Proof}. Since $g_2\leq q-q_1$, we can choose $g_2$ distinct
elements $b_1,\cdots, b_{g_2}$ from the set ${\bf F}_q - {\bf
F}_{q_1}$. For any element $\beta \in {\bf F}_{q^h}^* ={\bf
F}_{q_1^{mh}}^*$, since $\F_{q_1} [\alpha] = \F_{q_1^{mh}}$, we can
apply Theorem 2.2 to deduce that
$${\beta \over (\alpha+b_1)\cdots (\alpha+b_{g_2})}
=(\alpha+a_1)\cdots (\alpha+a_{g_1}),$$ where the $a_i\in {\bf
F}_{q_1}$ are distinct. The number of such sets $\{ a_1, a_2, a_3,
\cdots, a_{g_1}\} \subseteq \F_{q_1}$ is greater than
$$  {1 \over g_1!} \left({q_1^{g_1} -{g_1\choose 2}q_1^{g_1-1} \over q_1^{mh}-1}
-(1+{g_1\choose 2})(mh-1)^{g_1} q_1^{g_1/2} \right). $$ Since ${\bf
F}_{q_1}$ and its complement ${\bf F}_q -{\bf F}_{q_1}$ are
disjoint, it follows that
$$\beta =(\alpha+b_1)\cdots (\alpha+b_{g_2})(\alpha+a_1)\cdots
(\alpha+a_{g_1})$$ is a product of exactly $g_1+g_2$ distinct linear
factors of the form $\alpha+a$ with $a\in {\bf F}_q$. \hfill\qed

We now take $g_1 =\lfloor q^{1/2m}\rfloor =\lfloor \sqrt{q_1}
\rfloor$ and $g_2=\lfloor cq\rfloor -g_1$ in the above theorem.
Thus, $g_1+g_2 =\lfloor cq \rfloor $. We need $g_2$ satisfying the
inequalities
$$0\leq g_2 \leq q-q_1 =q -q^{1/m}.$$
That is,
$$0\leq \lfloor cq\rfloor -\lfloor q^{1/2m}\rfloor \leq q -q^{1/m}.$$
The left side inequality is satisfied if $q_1\geq c^{-2/(2m-1)}$.
The right side inequality is satisfied if $q_1\geq
(1-c)^{-1/(m-1)}$. Thus, we obtain

\begin{theorem} Let $m\geq 2$ and $h\geq 2$ be two positive
integers such that $q=q_1^m$. Let $0<c<1$ be a constant such that
$$q_1 \geq \max((mh-1)^{2+\epsilon}, ({4\over
\epsilon}+2)(mh+1)^2, c^{-2\over 2m-1}, (1-c)^{-1\over m-1})$$ for
some constant $\epsilon >0$. Then, every element in ${\bf
F}_{q^h}^*$ can be written as a product of exactly $\lfloor
cq\rfloor $ distinct linear factors of the form $\alpha+a$ with
$a\in {\bf F}_q$.
\end{theorem}

Combining this theorem together with Theorem 2.1, we deduce

\begin{theorem}
Let $m\geq 2$ and $h\geq 2$ be two positive integers such that
$q=q_1^m$. Let $0<c<1$ be a constant such that
$$q_1 \geq
\max((mh-1)^{2+\epsilon}, ({4\over \epsilon}+2)(mh+1)^2, c^{-2\over
2m-1}, (1-c)^{-1\over m-1})$$ for some constant $\epsilon >0$. Then,
the maximal likelihood decoding of the  Reed-Solomon code $RS_q[q,
\lfloor cq\rfloor -h]$ is at least as hard (in random time
$q^{O(1)}$ reduction) as the discrete logarithm in ${\bf
F}_{q^h}^*$.
\end{theorem}

Taking $m=2$ in this theorem, we deduce Theorem 1.1.

\begin{prop}

Let $h$ be a positive integer and
$0<c<1$ be a constant. Let $q_1$ be a prime power such that
\begin{equation} \label{equationh}
q_1 \geq
\max((2h-1)^{2+\epsilon}, ({4\over \epsilon}+2)(2h+1)^2, c^{-2/3},
(1-c)^{-1})
\end{equation}
for some constant $\epsilon >0$. Let  $q = q_1^2$. Let $h(x)$ be an
irreducible polynomial of degree $h$ over $\F_q$ whose root $\alpha$
satisfies that ${\bf F}_{q_1} [\alpha] = {\bf F}_{q^h}$. Let $f(x)$
be a nonzero polynomial over $\F_q$ of degree less than $h$. Then in
the Reed-Solomon code $RS_q[q, \lfloor cq\rfloor- h] $, the Hamming
ball centered at $({f(a)\over h(a)} + a^{\lfloor cq\rfloor
-h})_{a\in {\bf F}_q}$ of radius $q - \lfloor cq\rfloor $ contains
at least $ exp( \Theta(q)) $ many codewords.

\end{prop}

{\bf Proof:} The number of codewords in the ball is greater than
$${q_1^{ \lfloor \sqrt{q_1}\rfloor /2 } \over \lfloor \sqrt{q_1}\rfloor ! }{q-q_1 \choose \lfloor cq\rfloor
 - \sqrt{q_1} }, $$
which is greater than ${q-q_1 \choose \lfloor cq\rfloor - \sqrt{q_1}
} = exp(\Theta(q)) $. \hfill\qed

{\bf Proof of Theorem~\ref{manycodewords}}. Let $q$ to be the square
of the $i$-th prime power (listed in increasing order). Assume that
$i$ is large enough such that $\sqrt{q} \geq \max(c^{-2/3},
(1-c)^{-1}) $. We then let $\epsilon $ to be $1/ \log q $ and $h$ to
be the largest integer satisfying (\ref{equationh}). It remains to
find an irreducible polynomial of degree $h$ over ${\bf F}_q$, whose
root $\alpha$ satisfies that ${\bf F}_{q_1} [\alpha] = {\bf
F}_{q^h}$. Let $p$ be the characteristic of $\F_q$.
We can use $\alpha$ such that $  \F_p [\alpha] = \F_{q^h}$. We need
to find an irreducible polynomial of degree $h\log_p q$ over $\F_p$.
It can be done in time polynomial in $p$ and the degree
\cite{Shoup90a}. Then we factor the polynomial over $\F_q$ and take
any factor to be $h(x)$. As for $f(x)$, we may simply let  $f(x) =
1$. \hfill\qed

\section{Conclusion and future research}

In this paper, we show that the maximal likelihood decoding of the
Reed-Solomon code is at least as hard as the discrete
logarithm for any given information rate. In our result, we assumed
that the cardinality of the finite field is not a prime. While this
is not a problem in practical applications, e.g. $q=256$ is quite
popular,
it would be interesting to remove this restriction, that is,
allowing  prime finite fields as well.

Many important questions about decoding Reed-Solomon codes remain
open. For example, little is known about the exact list decoding
radius of  Reed-Solomon codes. In particular, does
there exist a Hamming ball of relative radius less than one which
contains super-polynomial many codewords in Reed-Solomon codes
of rate less than one?


\end{document}